\documentclass[twocolumn]{aastex701}
\usepackage{graphicx}
\usepackage{makecell}

\begin{document}
\newcommand{\htwo}{\ensuremath{\mathrm{H_2}}}

\title{A JWST NIRCam/MIRI view of the W51A high-mass star-forming region}

\author[0000-0003-2968-5333]{Taehwa Yoo}
\affiliation{Department of Astronomy, University of Florida, PO Box 112055, Florida, USA}
\email[show]{astro.taehwa.yoo@gmail.com}

\author[0000-0001-6431-9633]{Adam Ginsburg}
\affiliation{Department of Astronomy, University of Florida, PO Box 112055, Florida, USA}
\email{adamginsburg@ufl.edu}

\author[0000-0002-0533-8575]{Nazar Budaiev}
\affiliation{Department of Astronomy, University of Florida, PO Box 112055, Florida, USA}
\email{nbudaiev@ufl.edu}  

\author[0000-0003-1480-4643]{Roberto Galv\'an-Madrid}
\affiliation{Instituto de Radioastronom\'ia y Astrof\'isica, Universidad Nacional Aut\'onoma de M\'exico, Antigua Carretera a P\'atzcuaro 8701, Ex-Hda. San Jos\'e de la Huerta, Morelia, Michoac\'an, M\'exico C.P. 58089}
\email{robertogalvanmadrid@gmail.com}

\author[]{Aden Dawson}
\affiliation{Department of Astronomy, University of Florida, PO Box 112055, Florida, USA}
\email{adendawson@ufl.edu}

\author[0000-0002-1313-429X]{Savannah Gramze}
\affiliation{Department of Astronomy, University of Florida, PO Box 112055, Florida, USA}
\email{savannahgramze@ufl.edu}

\author[0000-0001-9797-5661]{Jes\'us Hern\'andez}
\affiliation{Universidad Nacional Aut\'{o}noma de M\'{e}xico, Instituto de Astronom\'{i}a, AP 106, Ensenada 22800, BC, M\'{e}xico}
\email{hernandj@astro.unam.mx}

\author[0000-0002-1379-4204]{Alexandre Roman-Lopes} 
\affiliation{Department of Astronomy, Universidad de La Serena, Av. Raul Bitran 1302, La Serena, Chile}
\email{aroman@userena.cl}

\author[0000-0001-8600-4798]{Carlos G. Rom\'{a}n-Z\'{u}\~{n}iga}
\affiliation{Universidad Nacional Aut\'{o}noma de M\'{e}xico, Instituto de Astronom\'{i}a, AP 106, Ensenada 22800, BC, M\'{e}xico}
\email{croman@astro.unam.mx}

\author[]{Joel Sanchez-Bermudez}
\affiliation{Universidad Nacional Aut\'onoma de M\'exico. Instituto de Astronom\'ia. A.P. 70-264, Ciudad de M\'exico, 04510, M\'exico\\}
\email{joelsb@astro.unam.mx}

\author[0000-0002-3941-0360]{Miriam G. Santa-Maria}
\affiliation{Department of Astronomy, University of Florida, PO Box 112055, Florida, USA} %\label{florida}
\affiliation{Instituto de Física Fundamental (CSIC). Calle Serrano 121-123, 28006, Madrid, Spain}
\email{miriam.g.sm@csic.es}

\author[0000-0001-8289-3428]{Aida Wofford}
\affiliation{Universidad Nacional Aut\'{o}noma de M\'{e}xico, Instituto de Astronom\'{i}a, AP 106, Ensenada 22800, BC, M\'{e}xico}
%\affiliation{Instituto de Radioastronom\'ia y Astrof\'isica, Universidad Nacional Aut\'onoma de M\'exico, Antigua Carretera a P\'atzcuaro 8701, Ex-Hda. San Jos\'e de la Huerta, Morelia, Michoac\'an, M\'exico C.P. 58089}
\email{awofford@astro.unam.mx}

\author[0000-0002-3576-4508]{Jason E. Ybarra}
\affiliation{Department of Physics and Astronomy, West Virginia University, 135 Willey Street, Morgantown, WV 26506, USA}
\email{jason.ybarra@mail.wvu.edu}

\begin{abstract}

We present observations of the W51A region, including the massive protoclusters W51-E and W51-IRS2, with JWST in 10 NIRCam and 5 MIRI filters.
In this work, we highlight the most novel features apparent in these images and compare them with other multi-wavelength images. The broad view of the NIRCam/MIRI images of the W51A region shows that areas dominated by warm dust and ionized gas are distinct from those dominated by PAHs. The high angular resolution of the JWST images resolves dust filaments in high contrast, revealing geometrically converging features feeding W51-E and a cavity around W51-IRS2. This picture adds support to the hypothesis that feedback from W51-IRS2 is suppressing further gas infall onto the protocluster, while by contrast, gas is still accreting onto W51-E.
Comparing the NIRCam and MIRI images to ALMA data, we find 24 sources detected by both JWST and ALMA, accounting for only $\sim10\%$ of the ALMA sources; the rest are too embedded or too cool to be detected by JWST. 
A knot of [Fe II] and H$_2$ emission north of W51-IRS2, previously detected in ground-based images, reveals peculiarly bright and compact peaks detected in all JWST bands.  The knot is likely the most energetic example of a protostellar jet driven by a massive star impacting dense interstellar medium. The new images provide a complementary view to the previous long-wavelength perspective on this 4 x 8 pc area of one of the most active star-forming regions in our Galaxy, revealing new mysteries to be further explored.

\end{abstract}


\section{Introduction}

The formation of massive stars is crucial for understanding how stellar feedback shapes the surrounding interstellar medium (ISM). 
In galaxies, a majority of the energy budget originates from massive stars ($M_*\gtrsim8\,M_\odot$) through various feedback processes, including ultraviolet radiation, stellar winds, and supernovae \citep[e.g.][]{zinnecker07, krumholz19b}

The W51 star-forming region stands out in our Galaxy because of its very active star formation compared to typical Galactic star-forming clouds. Among the three different sub-regions (A, B, and C), W51A is the youngest region \citep[$t<1\,{\rm Myr}$;][]{dawson25}, where massive stars are presently forming. In particular, the G49.5-0.4 component in W51A contains two protoclusters, W51-E (also known as W51 Main) and W51-IRS2, which are each potentially forming $>10^4\,\mathrm{M}_\odot$ of stars based on their current dense gas mass \citep[e.g.][]{ginsburg12}. As summarized in the review by \cite{ginsburg17b}, W51 has several observational advantages compared to similarly young, massive cluster-forming regions. It has relatively low line-of-sight dust extinction, particularly high radial velocity ($v\sim55\,{\rm km/s}$) at given Galactic longitude, and is relatively nearby ($d\sim5.4\,{\rm kpc}$; \citealt{sato10}), providing a unique opportunity to study high-mass star formation within high-mass clusters.

Throughout extensive observations across multiple wavelengths, W51A has revealed various signatures of high-mass star formation. In radio frequencies, following the first detection of free-free emission by \cite{westerhout58}, \cite{mehringer94} labeled several H\textsc{ii} regions ($\sim20$) in W51A using the Very Large Array (VLA). \cite{ginsburg16} utilized the Karl G. Jansky Very Large Array (JVLA) to better resolve the hyper/ultra-compact regions (HCH\textsc{ii}/UCH\textsc{ii} regions) in W51A and measure the outflowing material from W51-IRS2. 

At millimeter wavelength, dust continuum emission from compact sources with a wide range of sizes has been observed, including cores \citep{ginsburg17, motte22, ginsburg22, louvet24} and massive protostars ($r\sim500$--$1000\,{\rm AU}$) \citep[e.g.][]{zapata08, goddi20, tang22}. In particular, the recent ALMA long-baseline observation detected over 200 compact sources referred to as ``PPOs (Pre/Protostellar Objects)" in W51A, which are either actively forming stars or are expected to do so in the future \citep{yoo25}. Moreover, several hot cores with rich chemistry associated with massive protostars have been found in W51A \citep[e.g.][]{zhang97, goddi16, ginsburg17, bonfand24} with rich chemistry. These hot cores are common sites of maser emission from various molecules, including OH \citep{etoka12}, ${\rm H_2O}$ \citep{imai02,eisner02}, ${\rm CH_3OH}$ \citep{phillips05, etoka12}, SiO \citep{morita92, eisner02}, ${\rm NH_3}$ \citep{gaume93, goddi15}, and CS \citep{ginsburg19}. 

Due to high dust extinction in this region \citep[$A_V$ up to $\sim$5 mag;][]{dawson25}, the embedded young stellar population has been mainly studied through infrared observations. Using the Okayama Astrophysical System for Infrared Imaging and Spectroscopy (OASIS) on a ground-based telescope, \cite{okumura00} observed young OB stars and concluded that the star formation in W51 occurred very recently ($t\lesssim1\,{\rm Myr}$). \cite{kumar04} estimated the masses of young stellar populations in the G49.5-0.4 component of the W51A region as $\sim10^4\,\mathrm{M}_\odot$ using the 3.8m United Kingdom Infrared Telescope (UKIRT). W51A also has abundant embedded young stellar objects (YSOs) detected in survey programs using Spitzer and 2MASS \citep{kang09, saral17}. In particular, the giant H\textsc{ii} regions, IRS1/main and IRS2, which are the brightest components in the infrared view of W51A, have been subjects of great interest due to a plenty of embedded IR-bright sources, including young massive stars and UCH\textsc{ii} regions, with near-IR and mid-IR observations using NACO \citep{barbosa08, figueredo08}, T-ReCS \citep{figueredo08, barbosa16}, Gemini/NIRI \citep{figueredo08, barbosa08}, Gemini/NIFS \citep{barbosa22}, SOFI \citep{bik19}, SOFIA \citep{lim19}, and GTC EMIR \citep{dawson25}. 

In this paper, we present the first JWST NIRCam and MIRI view of the W51A region. The JWST NIRCam and MIRI imaging provide a sub-arcsecond resolution infrared view for the first time, revealing the detailed structures of dust and ionized gas in the high-mass star-forming region. Furthermore, individual embedded young stellar objects are resolved in the JWST imaging, enabling a more complete census of the stellar population in the W51A region. In future papers, we will characterize newly discovered compact sources along with the photometric SED fitting.

%The paper is organized as follows. In Sec.~\ref{sec:imaging}, we introduce our NIRCam/MIRI imaging data and its reduction. In Sec.~\ref{sec:results}, we report key features found in the NIRCam/MIRI images.  

\section{NIRCam and MIRI Imaging}
\label{sec:imaging}
\begin{figure*}
\centering
\includegraphics[scale=0.35]{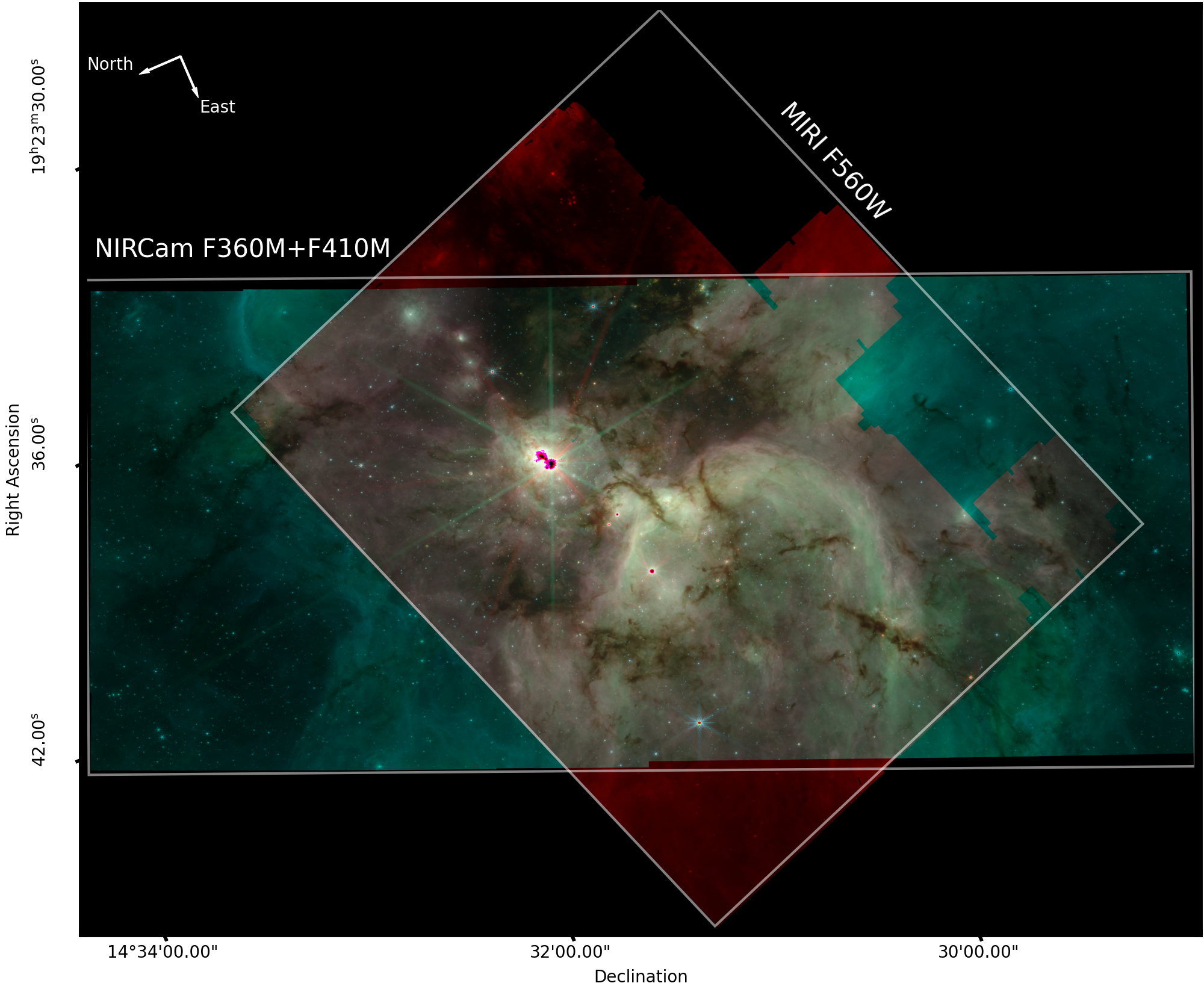}
\caption{An overview of W51A region. The composite image is produced by combining NIRCam F360M (blue), F410M (green), and MIRI F560W (red). The north and east directions in ICRS coordinates are marked as arrows at the upper left corner.}
\label{fig:footprint}
\end{figure*}

The JWST NIRCam and MIRI imaging of the W51A region were obtained on 2024 September 8 and 2025 May 6, respectively (program ID: 6151, PI: T. Yoo and A. Ginsburg). We summarize the NIRCam and MIRI observations in Table ~\ref{tab:obs}. The NIRCam observations were carried out with five filter combinations: (F162M+F150W2, F335M), (F182M, F360M), (F210M, F410M), (F140M, F480M), and (F187N, F405N+F444W) with an exposure time of 1890 seconds. The exposure time was calculated based on the YSO models from \cite{richardson24}, where the grids of YSO models with different geometries and view angles are provided. The YSO magnitudes at 20\% and 80\% of the total luminosity in each YSO geometry model were averaged over the geometry models and examined with the JWST Exposure Time Calculator (ETC) for saturation and detectability. We requested a MIRI imaging exposure time of 56 seconds, which is close to the minimum exposure time and was selected to minimize saturation. 

\begin{deluxetable}{ccc}
\label{tab:obs}
\tablecaption{Summary of NIRCam/MIRI observations}
\tablehead{& NIRCam & MIRI}
\startdata
Filters & \makecell{F140M, F162M,\\ F182M, F187N, \\F210M, F335M,\\F360M, F405N,\\F410M, F480M} & \makecell{F560W, F770W,\\F1000W, F1280W,\\F2100W} \\
Dither pattern & FULLBOX & \makecell{4-point Cycling\\Large} \\
Readout & SHALLOW2 & FASTR1 \\
Groups/Int & 5 & 5 \\
Integration/Exp & 1 & 1 \\
Exposure/Dither & 1 & 1 \\
Total dithers & 8 & 4 \\
Exposure time (s) & 1890 & 56 \\
\enddata

\end{deluxetable}

%Each NIRCam observation has a setup of FULLBOX primary dithers, SHALLOW2 readout pattern, Groups/Int=5, Integration/Exp=1, Subpixel dither type=SMALL-GRID-DITHER, small grid dither=2, resulting total of 8 dithers.
%For MIRI imaging, F560W, F770W, F1000W, F1280W, and F2100W filters are selected with 4-point Cycling/Large dithering,  Groups/Int=5, Integration/Exp=1, Exposures/Dither=1 with a total of 4 dithers. Each MIRI observation has a 56s exposure time.

%For the same reason, the FASTR1 readout pattern was chosen. 

The pointings of the NIRCam \citep{rieke23} and MIRI \citep{dicken24} observations were set to cover W51-E and W51-IRS2 protoclusters with a FULLBOX NIRCam primary dithering pattern and 2x2 full-array mosaic in MIRI. The footprints of the NIRCam and MIRI observations are shown in Fig.~\ref{fig:footprint}. 

The data reduction was conducted with the standard JWST pipeline Python package version 1.17.1 \citep{jwst_pipeline_1.17.1}. In the ramp fitting step of the stage 1 pipeline, we expanded the flux calculation to the pixels that are saturated in only one group by turning off \texttt{suppress\_one\_group}. This treatment allowed for a slight increase in the number of pixels available for flux estimation in the saturated region.
In the MIRI imaging, however, the saturated pixels are widely spread in the diffuse emission around W51-IRS2. We maximized the available pixels by removing the \texttt{\textsc{DO\_NOT\_USE}} flags from the first frame and for all saturated pixels. We took advantage of the very high flux of saturated pixels compared to the transient noise in the first frame because the relative uncertainty is not relatively large. Despite this treatment, the MIRI images at long wavelengths, e.g., F1000W, F1280W, and F2100W, are still severely saturated over large areas around W51-IRS2 and W51 IRS1/main shell. We defer the recovery of the MIRI saturated pixels and the detailed MIRI imaging analysis to future work.

 For the NIRcam images, 1/f noise was removed in post-processing using the \texttt{image1overf} method\footnote{https://github.com/chriswillott/jwst/blob/master/image1overf.py}.

 All the JWST data used in this paper can be found in MAST: \dataset[https://doi.org/10.17909/ax7r-bw30]{https://doi.org/10.17909/ax7r-bw30}.

\section{Results}
\label{sec:results}

In this section, we highlight a new JWST view of the region, including dust ridge structures, H\textsc{ii} regions, outflows, and protoclusters. We also provide an overview of the JWST compact sources that are spatially associated with a previous ALMA compact source catalog. We characterize objects in this work by their apparent properties, such as morphology; detailed quantitative analysis will be performed in future work.

%\begin{figure*}
%    \centering
%    \includegraphics[scale=0.7]{paabra_ratio.png}
%    \caption{PaA/BrA map of W51A region. The ratio is obtained by dividing F187N image by F405N image. The horizontal noise on both edges of the image is due to 1/f noise in the F187N image. The white pixels on the right bottom corner are created by bad pixels in the F405N image.}
%    \label{fig:paabra}
%\end{figure*}

\subsection{Large-scale structures}

\begin{figure*}
    \centering
    \includegraphics[scale=0.8]{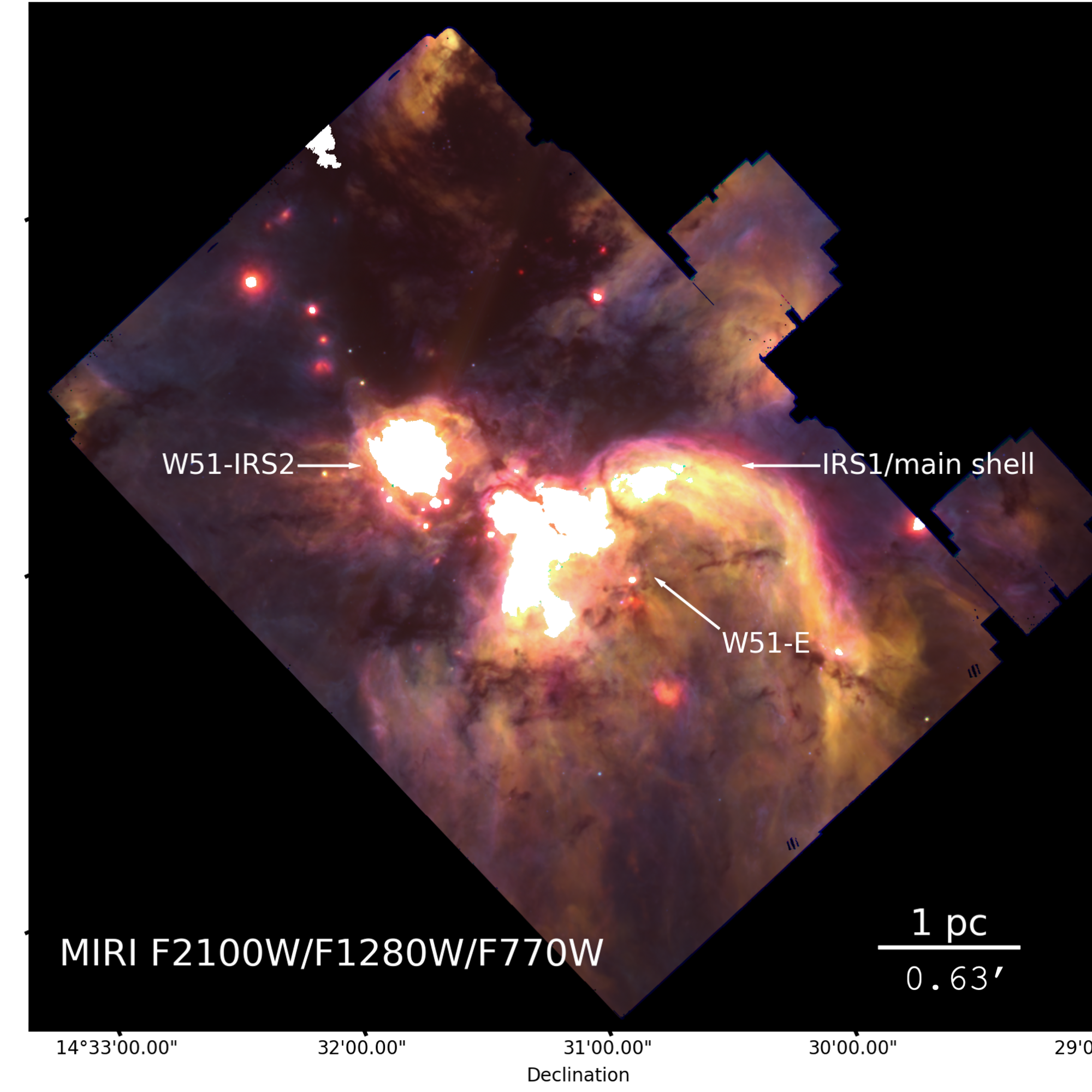}
    \caption{W51A MIRI overview. The image shows a composite of F770W (blue), F1280W (green), and F2100W (red). The pixels saturated in any filters are filled with white color. }
    \label{fig:miri}
\end{figure*}

\begin{figure*}
\centering
\includegraphics[scale=0.6]{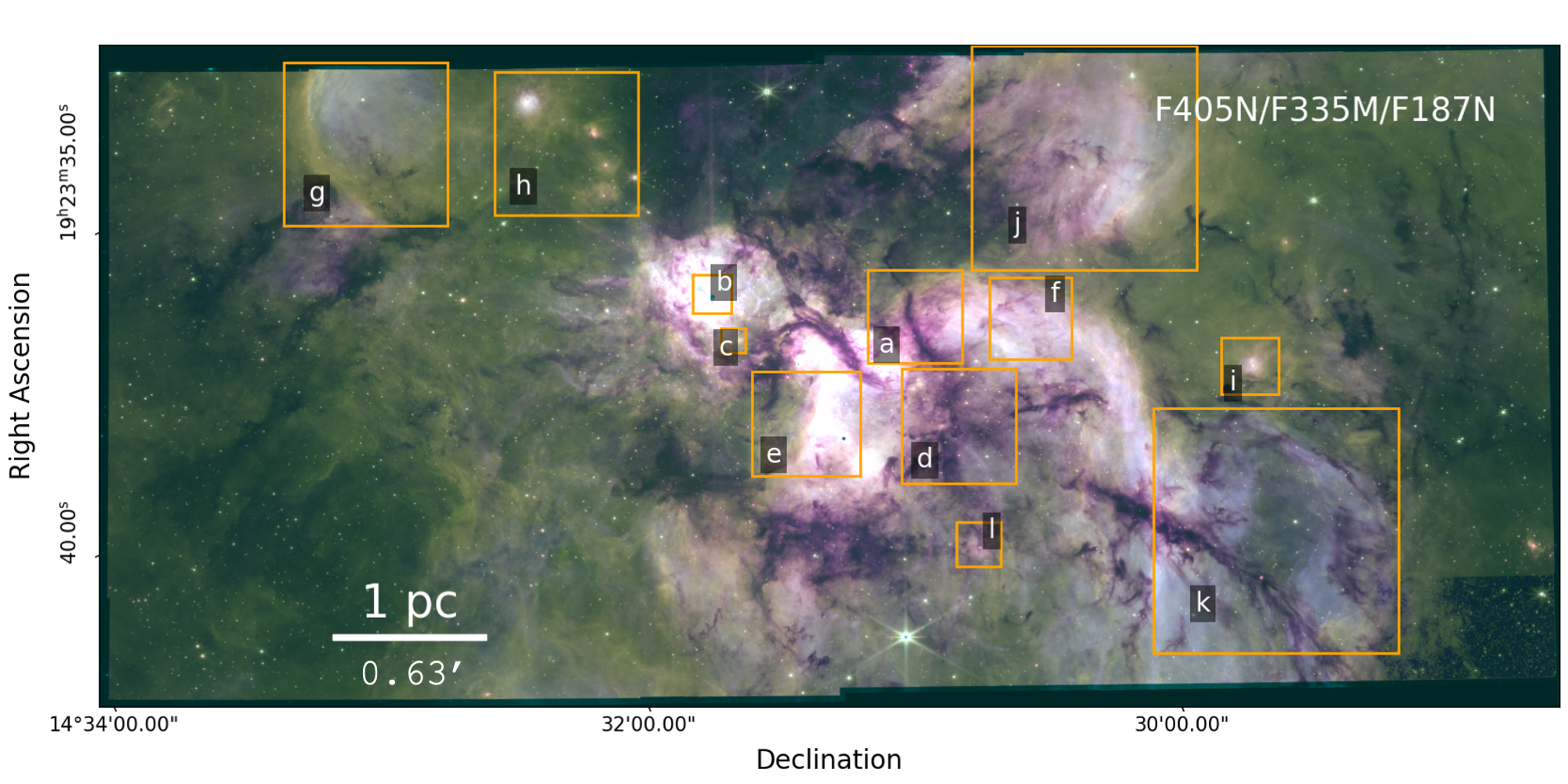}
\caption{W51A NIRCam overview showing a F405N (red)/F335M (green)/F187N (blue) composite image. The positions of the cutouts for the zoomed-in images displayed in Fig.~\ref{fig:higlights_cutouts} are marked.}
\label{fig:highlight1}
\end{figure*}

\begin{figure*}
    \centering
    \includegraphics[scale=0.26]{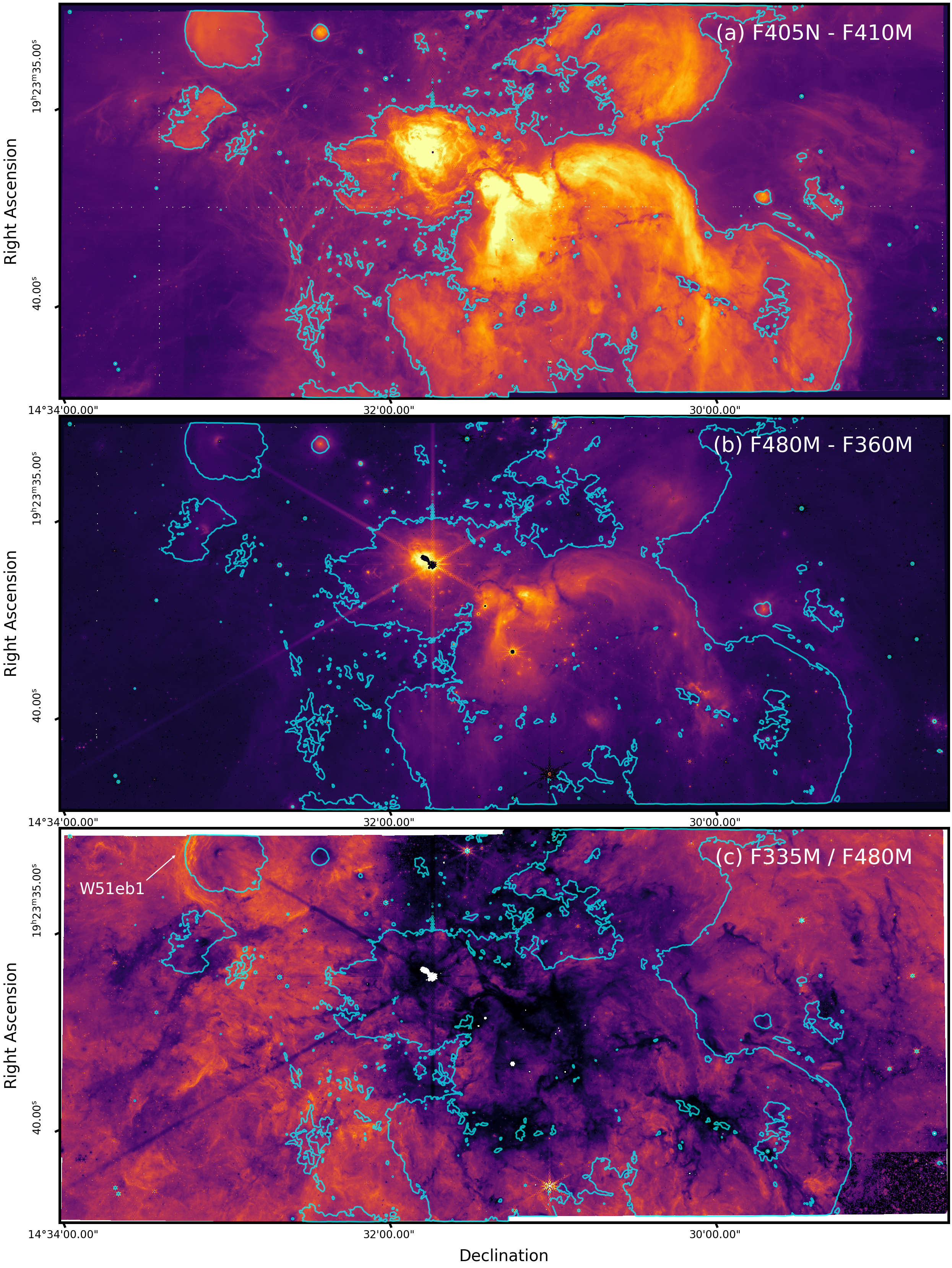}
    \caption{Multiple color images of W51A region in a large-scale view. The images of F405N-F410M (top), F480M-F360M (middle), and F335M/F480M (bottom) are displayed to represent ionized gas, warm dust, and PAH abundance, respectively. The cyan contour marks the flux density of F405N-F410M image at 250 MJy/sr. }
    \label{fig:multiple_view}
\end{figure*}

\begin{figure*}
\centering
\includegraphics[scale=0.47]{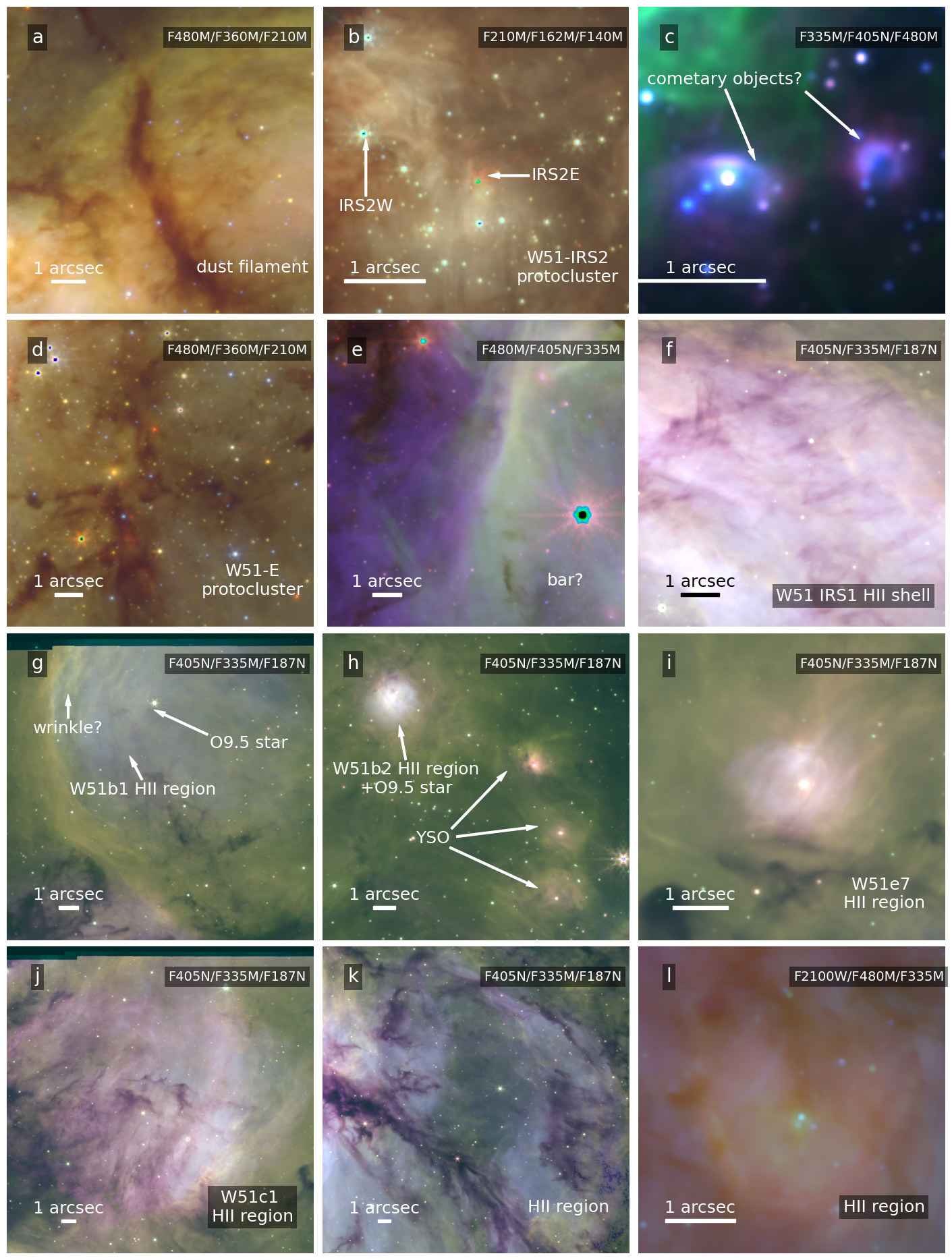}
\caption{The cutout images of the highlighted boxes in Fig.~\ref{fig:highlight1}. \textbf{a:} A dust filament around W51-E. \textbf{b:} W51-IRS2 protocluster. \textbf{c:} Cometary objects around W51-IRS2. \textbf{d:} W51-E protocluster. \textbf{e:} A bar at the edge of IRS1 H\textsc{ii} region. \textbf{f:} W51 IRS1 H\textsc{ii} region shell structure. \textbf{g:} W51b1 H\textsc{ii} region. \textbf{h:} W51b2 H\textsc{ii} region and YSOs. \textbf{i:} W51e7 H\textsc{ii} region. \textbf{j:} W51c1 H\textsc{ii} region. \textbf{k and l:} Newly discovered H\textsc{ii} regions. }
\label{fig:higlights_cutouts}
\end{figure*}

We present the MIRI and NIRCam overviews of the W51A region in Figs.~\ref{fig:miri} and \ref{fig:highlight1}. %\textbf{We also provide the details of several highlighted features in Fig.~\ref{fig:higlights_cutouts}.} 
The JWST images are particularly focused on the G49.4-0.5 component of the W51A region that was first observed as two bright main components, IRS1 and IRS2, in the infrared observations of \cite{wynn-williams74}. IRS1 is a infrared-bright spot in the middle of the arc surrounding the W51-E/main protocluster region, and IRS2 is the place where W51-IRS2 protocluster is found at millimeter wavelength but also bright in infrared.
We will first provide a large-scale view of W51A region using different colors to provide an insight into which component is responsible for the NIRCam/MIRI emission. 
We then highlight dust filaments and protoclusters in this section and will return to the H\textsc{ii} regions in Sec.~\ref{subsec:hii}.
%The details of this giant H\textsc{ii} region will be discussed in Sec.~\ref{subsec:H\textsc{ii}}. Here, we feature dust filaments and protocluster regions in the JWST image. 

\textit{W51A region with different colors}---In Fig.~\ref{fig:multiple_view}a, we create an F405N-F410M image to trace Br $\alpha$ recombination line from ionized gas. Several H\textsc{ii} regions, including IRS1, IRS2, W51b1, and W51c1 (Sec.~\ref{subsec:hii}) are well highlighted. We also display a F480M-F360M difference image (Fig.~\ref{fig:multiple_view}b) to trace the warm dust component, e.g., $T=500\,{\rm K}$, which peaks at $\lambda\sim6\,{\rm \mu m}$. 
In most cases, the F480M-F360M is bright towards H\textsc{ii} regions, particularly the inner region around the driving source of the H\textsc{ii} region, which heats the dust inside the H\textsc{ii} regions. The overall overlap between warm dust and ionized gas indicates the existence of dust grains inside the ionized region \citep[e.g.][]{povich07, paladini12, ochsendorf15}.

In particular, the dust emission within H \textsc{ii} regions has often been attributed to the stochastic heating of very small grains (VSGs), which are typically detected around $\lambda \sim 20,{\rm \mu m}$ \citep[e.g.,][]{paladini12, anderson12}. In the MIRI view of the W51A region (Fig.~\ref{fig:miri}), several H \textsc{ii} regions exhibit bright F2100W emission, including the IRS1/main arc and compact, centrally concentrated regions such as W51b2 (inset h of Fig.~\ref{fig:highlight1}) and the H \textsc{ii} region shown in inset l of Fig.~\ref{fig:highlight1}, which will be discussed later. Although direct evidence for very small dust grains is still lacking, the compact F2100W emission seen in these H \textsc{ii} regions may suggest their presence, as such small grains are less efficiently expelled by radiation pressure compared to larger grains \citep{draine11, akimkin15}.

On the other hand, the F335M/F480M ratio shows a completely different structure (Fig.~\ref{fig:multiple_view}c). In the case of the Orion bar, 50-60\% of emission detected in the F335M filter comes from PAH emission \citep{chown24}. Using this as a benchmark, the ratio with respect to F480M, which does not cover any PAH features, serves as an indicator of PAH excitation and destruction. This is similar to the indicator [5.8]/[4.5] that was previously used for tracing PAH destruction in \cite{povich07}. In Fig.~\ref{fig:multiple_view}c, the F335M/F480M emission appears faint in regions that are bright in F480M–F360M and F405N-F410M, i.e., H\textsc{ii} regions. This reconfirms previous reports that PAHs are mainly destroyed in H\textsc{ii} regions \citep[e.g.,][]{povich07, egorov23}. On the other hand, the regions bright in F335M/F480M mark PDRs where FUV excites PAH emission. In particular, W51b1, the H\textsc{ii} region at the upper left corner, has a prominent F335M/F480M excess around the boundary.

\textit{Dust filaments}---There are several dark lanes crossing in front of the IRS1 H\textsc{ii} shell in both MIRI and NIRCam images (Fig.~\ref{fig:miri} and \ref{fig:highlight1}). These lanes have been identified in multiple previous observations at infrared wavelengths, but their detailed morphology had not been clearly resolved \citep[e.g.][]{goldader94,lim19}.
The dark lanes, showing absorption by cold dust filaments against the bright background emission (Fig.~\ref{fig:highlight1} and Fig.~\ref{fig:higlights_cutouts}a), exhibit numerous dendritic structures.
%The morphology in the NIRCam image (Fig.~\ref{fig:highlight1}a and Fig.~\ref{fig:higlights_cutouts}a) reveals numerous dendritic structures, directly telling us that the dark lanes are the absorption feature of the cold and dust filaments against the bright background emission.
In the inset Fig.~\ref{fig:highlight1}d, several dust filaments appear to converge at W51-E, hinting that material is still actively inflowing onto the protocluster. However, additional kinematic studies are required to confirm this scenario.

 \textit{W51-E and W51-IRS2 protoclusters}---The two protoclusters W51-E and W51-IRS2 are actively star-forming regions that host densely populated UCH\textsc{ii} and HCH\textsc{ii} regions \citep[e.g.][]{gaume93,  mehringer94, zhang97, ginsburg16} and dust continuum sources \citep[e.g.][]{zapata08, goddi20, tang22, louvet24,  yoo25}. The W51-E protocluster is located where the filaments converge (marked in inset  \ref{fig:highlight1}d). In contrast, no dust filaments are seen at the inner place of the W51-IRS2 protocluster (around inset b in Fig.~\ref{fig:highlight1}) - if they existed, they are now cleared out. This clearing is consistent with the previously suggested hypothesis that the feedback from W51-IRS2 suppresses mass infall onto the protocluster \citep{ginsburg16}. The inner region of the W51-IRS2 protocluster is more clearly shown in panel b of Fig.~\ref{fig:higlights_cutouts}. Many compact sources are found at shorter wavelengths $\lambda\lesssim2\,\mu {\rm m}$, whereas longer wavelength images suffer from saturation due to very bright emission from IRS2E. In Fig.~\ref{fig:higlights_cutouts}c, two cometary objects are identified. These objects might be either evaporating gas globules \citep{sahai12} or proplyds where the shapes of the cometary gas are facing toward W51-IRS2. Along with the geometry of dust filaments around W51-IRS2, this may show the impact of feedback from W51-IRS2. In contrast, the W51-E protocluster in Fig.~\ref{fig:higlights_cutouts}d shows dust filaments apparently intruding into the inner region.

\subsection{H\textsc{ii} regions}

\label{subsec:hii}
The W51A overview images in Figs.~\ref{fig:miri}, ~\ref{fig:highlight1}, and \ref{fig:multiple_view} reveal the IRS1/main H\textsc{ii} region arc around W51-E protocluster and several different UCH\textsc{ii} regions with different morphologies and colors. The IRS1/main arc is an H\textsc{ii} region that has a diameter of $\sim1\,{\rm arcmin}$ corresponding to $1.5\,{\rm pc}$ at a distance of W51A, $d=5.4\,{\rm kpc}$. In both NIRCam and MIRI image, the lower (approximately east in ICRS coordinates) half boundary of the H\textsc{ii} region is not clearly defined, but the tails at both ends of the arc run in parallel, indicating that this is a blister H\textsc{ii} region that sweeps out the ambient medium. 

The multi-color images in MIRI and NIRCam provide a detailed view of the structure of the arc. The arc shows substructures with three or more ridges in F335M (panel f of Fig.~\ref{fig:higlights_cutouts}) that were identified in a previous VLA observation \citep{ginsburg16}. The striping pattern of F335M is likely due to different FUV optical depth, resulting in different PAH destruction levels.

%The peak positions of the ridges have offsets between F335M and F405N, which manifests as an alternating stripe pattern. The offset can be attributed to different gas density across the edge of the IRS1/main, leading to different optical depth for FUV to destroy the PAH beyond \citep{habart24}. 

%not sure if I put this paragraph into paper or not. This paragram consume a lot of space for just artifact and the explanation is not clear
%The arc is also bright in MIRI filters with similar substructure. The outer edge is typically bright in F2100W, but it could be an artifact of severe saturation. We turned off \texttt{first\_frame} and \texttt{saturation} in the MIRI stage 1 pipeline to maximize available pixels but the flux estimation derived from the slope fitting for saturated pixel ramps has limitation in particular for pixels saturated at their first frame. As a result, the inner region of IRS1 likely has underestimated pixel values, which in turn makes the outer edge appear relatively brighter in F2100W.

At approximately the northern side (left side in Fig.~\ref{fig:highlight1}) of the arc, the edge has a bar shape (panel e of Fig.~\ref{fig:higlights_cutouts}), which is analogous to the Orion bar \citep{habart24}. PAH-dominated areas and ionized regions are well-separated by the bar. 

There are UCH\textsc{ii} regions already studied in \cite{barbosa22} where Spitzer NIR compact sources \citep{barbosa16} are matched with UCH\textsc{ii} regions observed in VLA \citep{mehringer94}. Among labeled UCH\textsc{ii} regions in \cite{mehringer94} and \cite{barbosa22}, W51b1 (Fig \ref{fig:higlights_cutouts}g), W51b2 (Fig \ref{fig:higlights_cutouts}h), W51e7 (Fig \ref{fig:higlights_cutouts}i), and W51c1 (Fig \ref{fig:higlights_cutouts}j) are discernible in the JWST image. 

W51b1 is the H\textsc{ii} region in Fig.~\ref{fig:higlights_cutouts}g. The driving source of W51b1 was characterized as an O9.5 star G09.4945-00.4329 \citep{saral17}. In the F405N/F335M/F187N image in Fig.~\ref{fig:higlights_cutouts}, the left edge of the H\textsc{ii} region is clearly visible, while the other edge is not clear. 
%Despite a similar blister-type shape as the IRS1/main arc, ionized gas  relatively evenly fills the inside of the H\textsc{ii} region. 
Near the upper left boundary, wrinkle patterns in F335M are observed. The notable characteristic of this H\textsc{ii} region is that its color is bluer (higher F187N/F405N) than that of other regions. This is probably due to either more dust scattering at short wavelengths or less dust extinction along the line-of-sight, which would place it further in the foreground than other structures.

We compare several of the compact H\textsc{ii} regions that exhibit different morphological features.
W51b2 is seen in the upper left of Fig.~\ref{fig:higlights_cutouts}h with a closed boundary as opposed to the blister-type H\textsc{ii} regions like W51b1. \cite{barbosa22} identified the star at the center of W51b2 as an O9.5 using NIR spectroscopy.
W51e7  has aligned streaks in its surrounding dust material (Fig.~\ref{fig:higlights_cutouts}i). The spectral type of the central star was classified as a B1 star in \cite{bik19} and an O9.5 star in \cite{barbosa22}. While W51b2 and W51e7 are embedded with a closed boundary (they are circularly symmetric), W51c1 has a more open morphology with a less clear boundary (Fig.~\ref{fig:higlights_cutouts}j). It is also likely to be a blister-type H\textsc{ii} region where its boundary is open to the relatively low-density side \citep{krumholz09}. 

We identified two possibly new H\textsc{ii} region candidates. Panel k of Fig.~\ref{fig:higlights_cutouts} shows an arc shape illuminated by recombination lines. Similar to W51eb1, F187N is more dominant over the region than F405N. The geometry of flocculent dust lanes approximately follows that of the H\textsc{ii} region boundary. Another candidate is found in Fig.~\ref{fig:higlights_cutouts}l. Unlike W51b1, F405N is dominant over F187N. This region is bright in other long wavelengths, especially in F2100W. % speculation: if it's affected by the ionizing gas flow from IRS1, then the grain size is much smaller than others? 

\subsection{ALMA-matched sources}

\begin{figure*}
    \centering
    \includegraphics[scale=0.55]{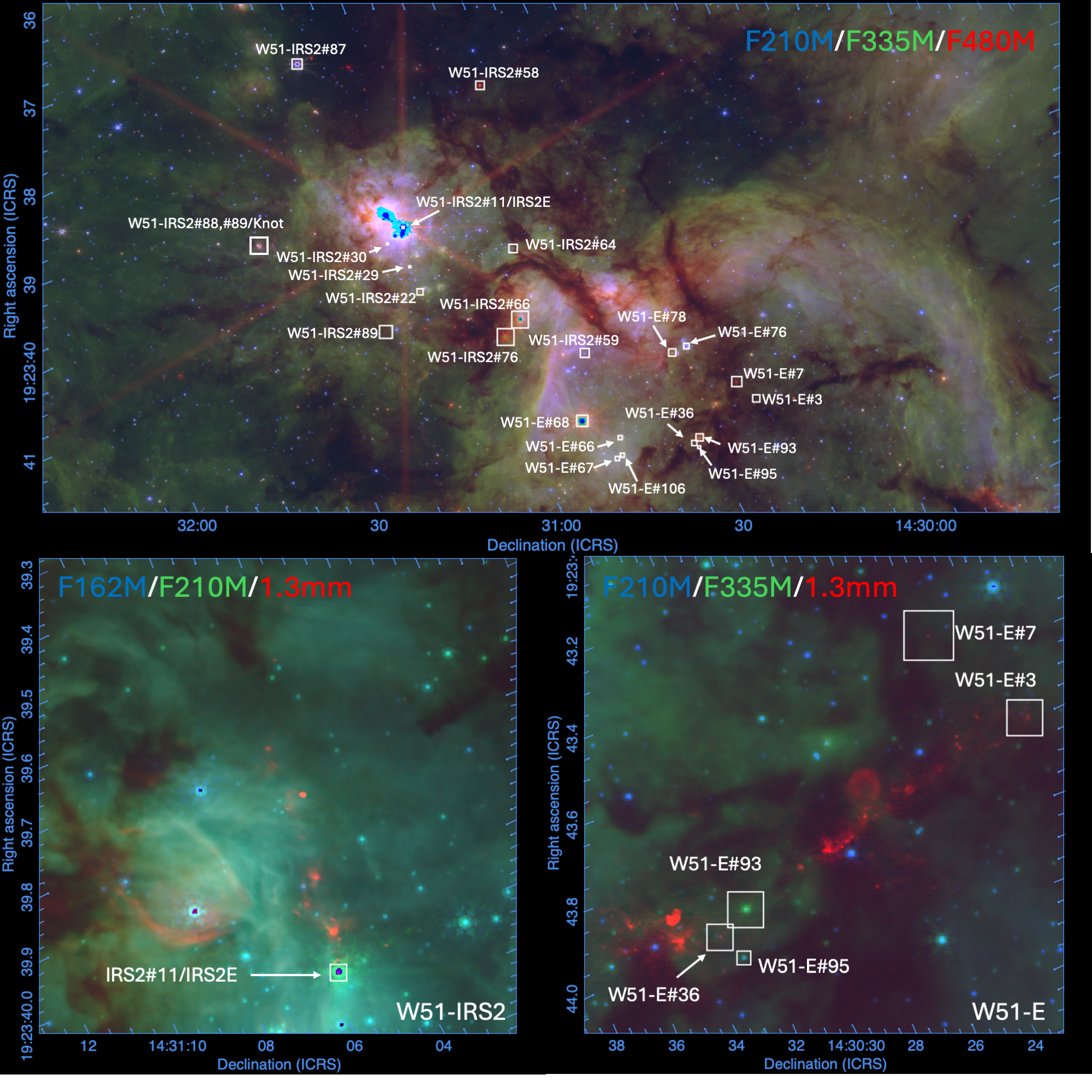}
    \caption{ALMA-JWST matched sources. The locations of the matching sources are marked in the upper panel with the background image of F162M, F210M, and F480M filters. In the lower panels, W51-E and W51-IRS2 protocluster regions are zoomed in with the background image of the JWST NIRCam filters and ALMA 1.3mm image combined. }
    \label{fig:alma_overlapping_map}
\end{figure*}

\begin{figure*}
\centering
\includegraphics[scale=0.35]{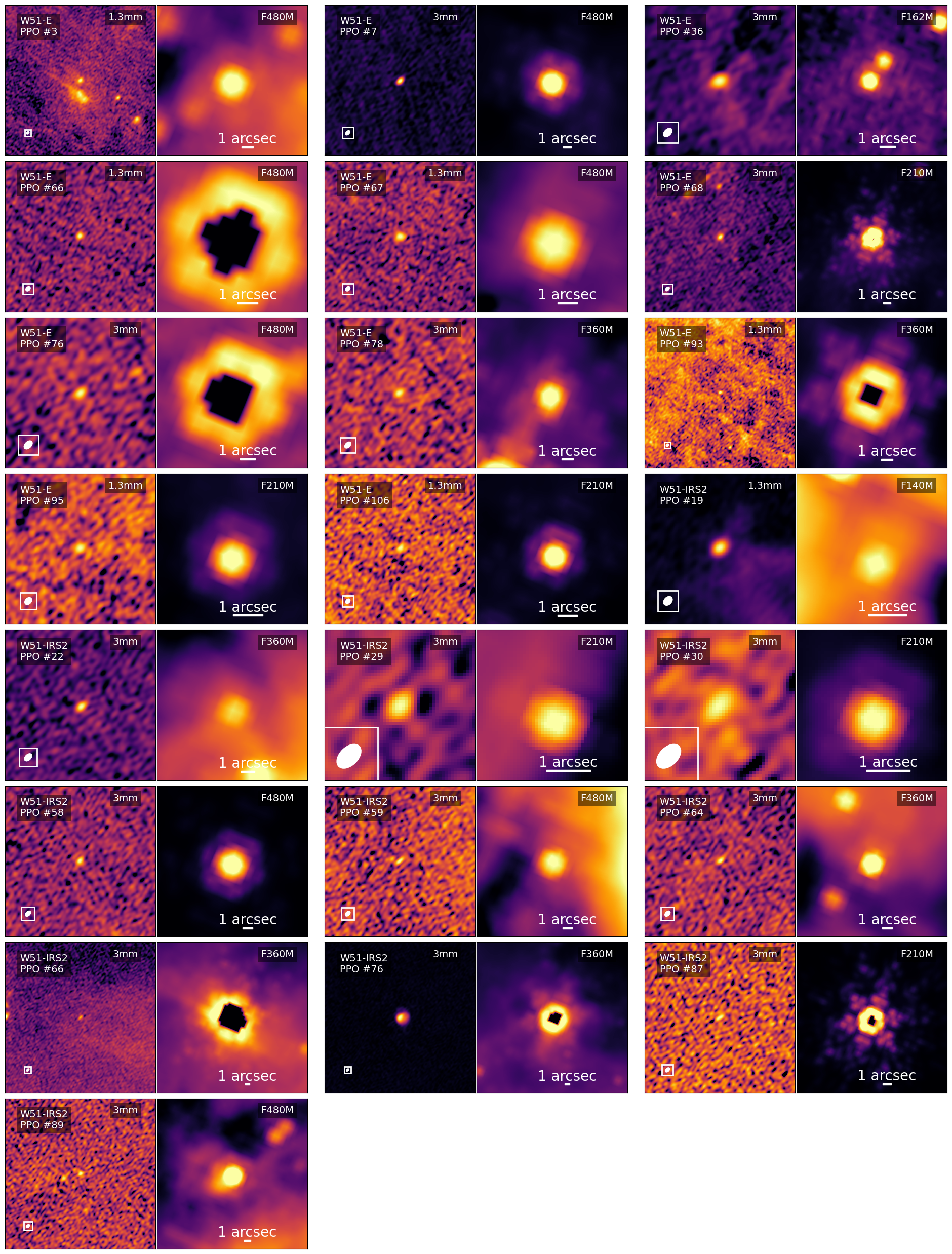}
\caption{ALMA and JWST cutout images for the ALMA PPOs that have a JWST counterpart in Fig.~\ref{fig:alma_overlapping_map}. The color stretch of each cutout image is adjusted to visually represent the source clearly. The synthesized beam size (0.027 arcsec at $1.3\,{\rm mm}$; 0.066 arcsec at $3\,{\rm mm}$) of the ALMA images is added to estimate the size of the cutout image. Some JWST PSFs have empty pixels at the center due to saturation. Note that the systematic offset $\sim0.04\,{\rm arcsec}$ between the JWST and ALMA image is uncorrected here. The W51-IRS2 PPO \#90 and \#91 are excluded in this figure as they will be covered in Fig.~\ref{fig:outflow}. }
\label{fig:alma_overlapping}
\end{figure*}

Previous ALMA high-resolution images detected 118 and 93 compact sources in W51-E and W51-IRS2, respectively \citep{yoo25}. The compact sources were referred to as ``PPOs (Pre/Protostellar Objects)" to bracket all different stages of star formation, prestellar/protostellar dust envelopes/disks, or hyper-compact H\textsc{ii} regions. 
% JH: I just added "the", removed a comma, and added a plural for clarity. 
We cross-matched these sources with NIRCam and MIRI point sources and found 11 and 13 matches in the W51-E and W51-IRS2 regions, respectively. Note that we found a uniform coordinate offset, $\sim0.04\,{\rm arcsec}$ between the ALMA continuum images and the JWST images. We take into account this offset when we match the sources. We show the locations of the matched sources in Fig.~\ref{fig:alma_overlapping_map} and their snapshots in Fig.~\ref{fig:alma_overlapping}. In Fig.~\ref{fig:alma_overlapping}, only 22 out of 24 matched sources are presented. The two excluded sources are the ALMA counterparts of the outflow knot in W51-IRS2, W51-IRS2 PPO \#88 and \#89, which will be covered in Sec.~\ref{subsubsec:knot}. 

Only 11\% (24 out of 211) of PPOs have JWST NIRCam or MIRI counterparts. In other words, only a small fraction of PPOs have low enough dust extinction along the line-of-sight to be seen in JWST NIRCam and MIRI. The matched sources are results of either a less embedded evolutionary phase compared to the non-matched sources or their relatively low ambient density, e.g., the outer region of a cloud. 
%A low occurrence of overlap between JWST and ALMA point sources suggests that ALMA traces the early stage of star formation, usually embedded in a dust envelope, whereas JWST detects a later stage where the dust envelope is relatively disrupted. 

When we cross-matched these sources with SIMBAD, we found 3 H\textsc{ii} regions (W51-E PPO\#76, \#78, W51-IRS2 PPO\#76), 1 YSO candidate (W51-E PPO\#68; \citealt{saral17}), and 1 AGB candidate (W51-IRS2 PPO\#87; \citealt{saral17, lim19}). The three H\textsc{ii} regions correspond to W51e20, e12, and e5 H\textsc{ii} regions, respectively, which are identified in \cite{ginsburg16} and \cite{rivera-soto20}. Among the unmatched sources, W51-E PPO\#7 has a measurement of spectral index between $1.3\,{\rm mm}$ and $3\,{\rm mm}$, $\alpha=0.58$ \citep{yoo25}, indicating it is likely a free-free emission source. The characterization of other matched sources requires a more complete spectral energy distribution analysis that we defer to future studies.  

\subsubsection{W51-IRS2E}

W51-IRS2E is a very bright source that saturates diffuse emission at $\gtrsim2\,{\rm \mu m}$ around the W51-IRS2 region. As summarized in \cite{ginsburg17b}, W51-IRS2E was observed in earlier NIR photometric and spectroscopic observations in \cite{figueredo08} and \cite{barbosa08}. We confirmed that this source has an ALMA compact source counterpart that was identified as W51-IRS2 \#19 in \cite{yoo25} (Fig.~\ref{fig:alma_overlapping}), also labeled as CA1 in \cite{tang22}. 

Previous NIR spectroscopy has revealed that W51-IRS2E shows CO overtone emission features at 2.3--$2.4\,{\rm \mu m}$ \citep{barbosa08}, suggesting the presence of a hot accretion disk of YSO. The ALMA counterpart is unresolved at both $1.3\,{\rm mm}$ and $3\,{\rm mm}$ (Fig.~32 of \citealt{yoo25}), and thus the size of the putative disk should be smaller than $\sim150$--$200\,{\rm AU}$. Notably, W51-IRS2E is spatially associated with a prominent X-ray ACIS source from the Massive star-forming regions Omnibus X-ray Catalog (MOXC; \citealt{townsley14}). \cite{townsley14} reported that the X-ray emission from W51-IRS2E is uncommonly hard, so much so that it produces a fluorescent Fe line at $6.5\,{\rm keV}$. 

Interestingly, no outflow feature has been observed despite the fact that the CO overtone and the possible shock-generating X-ray emission may indicate the presence of accretion. This is in contrast to the neighboring massive protostar, W51north, producing molecular outflow \citep{goddi20} with a vast amount of diffuse dust emission around it. We will discuss this outflow feature further in the next section. Another possible cause of X-ray emission is colliding wind binaries, although its companion has not been found. 
%The apparently closest neighbor confirmed so far is W51-E PPO \#20, which is 0.4 arcsec away from W51-IRS2E. 

\subsection{Molecular hydrogen shocked emissions and outflows}

%\begin{figure*}
%    \centering
%    \includegraphics[scale=0.4]{cutout_h2k_jwst_newnew.png}
%    \caption{Cutouts of \htwo\ emission in GTC observation of \citep{dawson25} and JWST observation. For each source, GTC (left) and JWST (right) images are shown for comparison. For sources outside the NIRCam field of view, the F560W image is shown instead of F210M.}
%    \label{fig:h2k}
%\end{figure*}
\begin{figure*}
    \centering
    \includegraphics[scale=0.47]{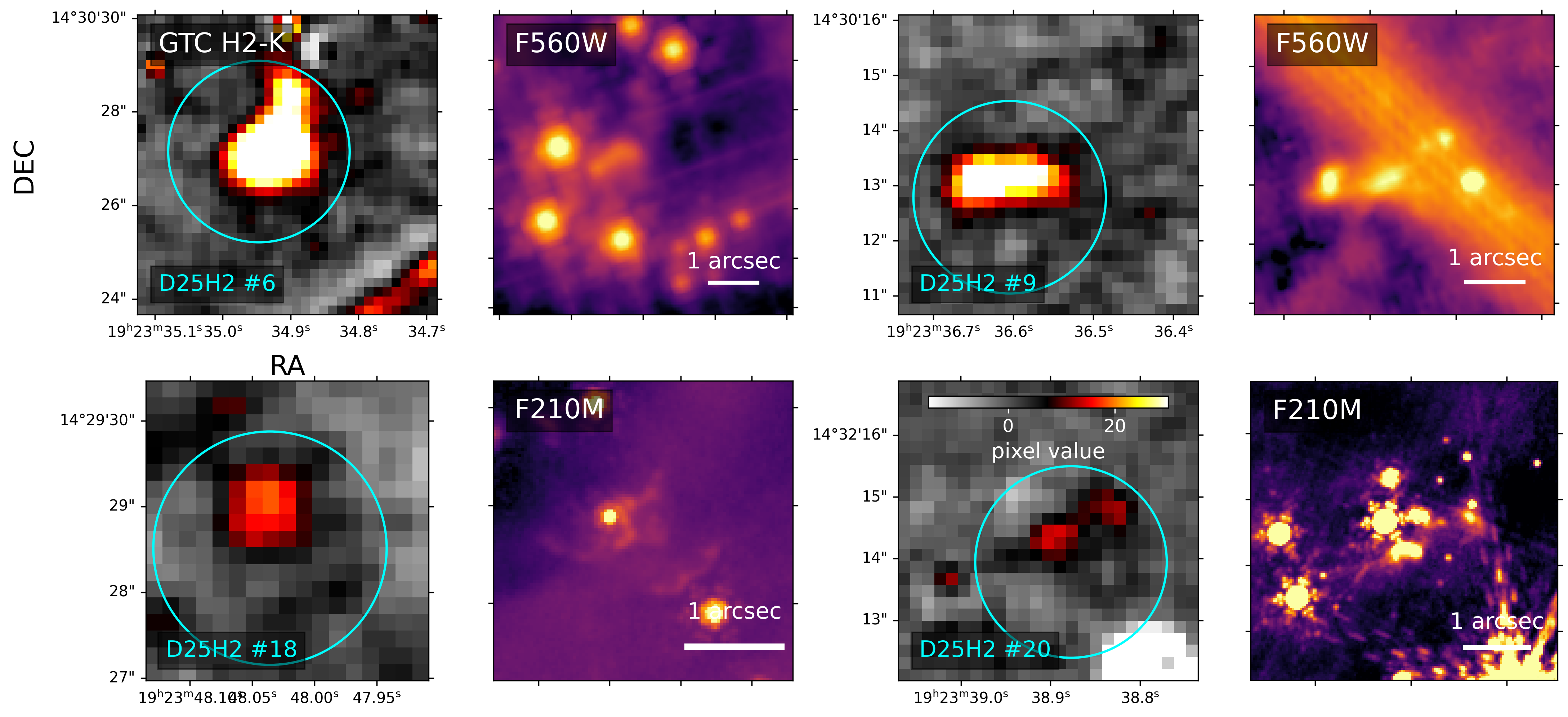}
    \caption{Cutouts of \htwo\ emission in GTC observation of \citep{dawson25} and JWST observation. For each source, GTC (left) and JWST (right) images are shown for comparison. For sources outside the NIRCam field of view, the F560W image is shown instead of F210M.}
    \label{fig:h2k}
\end{figure*}
We compare our JWST images with previously observed GTC images in an  H2 narrow-band filter and a K-band filter. Molecular hydrogen can be excited either by UV radiation or shocks. Therefore, \htwo\ excess indicates the presence of UV-producing massive stars or protostellar outflows. The \htwo\ emission in W51A region has been observed in ground-based telescopes \citep{hodapp02,dawson25}.

In this section, we attempt to better characterize \htwo\ emission features using the higher angular resolution of our JWST images.  Most of the GTC objects lie within the field of view of the JWST NIRCam observations, which include the F210M filter that covers the H2 S(1) 1-0 2.12$\mu {\rm m}$ line that was observed by the GTC.  In the case of \htwo\ emission found outside of our NIRCam field of view, we instead present our MIRI F560W images. F560W emission is not primarily from \htwo\ emission, but it can reveal either driving YSOs or shock-heated dust.

In the work of \cite{dawson25}, the K-band image was subtracted from the H$_2$ narrowband image to produce continuum-subtracted H$_2$ images.  In these images, objects with excess emission were selected by eye. Hereafter, we will use D25H2 to label the \htwo\ emission features cataloged in Fig.~12 of \cite{dawson25}; the numbers used in Fig.~\ref{fig:h2k} correspond to this catalog. Only 5 of the 21 \htwo\ excess sources reported by \citet{dawson25} were better-resolved in the JWST images; other \htwo\ excess objects are not detected in the JWST images, which may indicate that they are spurious detections produced by the difference imaging approach or that the more selective narrow-band filter is better able to select for H$_2$ emission in regions with bright, extended K-band continuum emission.

In Fig.~\ref{fig:h2k}, 4 of these 5 sources are shown; the other source cataloged as D25H2 \#1 will be covered in detail in Sec.~\ref{subsubsec:knot}. The JWST counterparts of D25H2 \#6 and \#9 in Fig.~\ref{fig:h2k} exhibit elongated features in F560W, suggesting that \htwo\ emission is associated with outflow. The D25H2 \#18 and \#20 show bow-like features in F210M, which is indicative of shocks. The comparison of GTC H2-K image with JWST images confirms that the some of excesses in the GTC H2-K image are associated with outflows.

\subsubsection{Knot and outflow in W51-IRS2}
\label{subsubsec:knot}
\begin{figure*}
    \centering
    \includegraphics[scale=0.9]{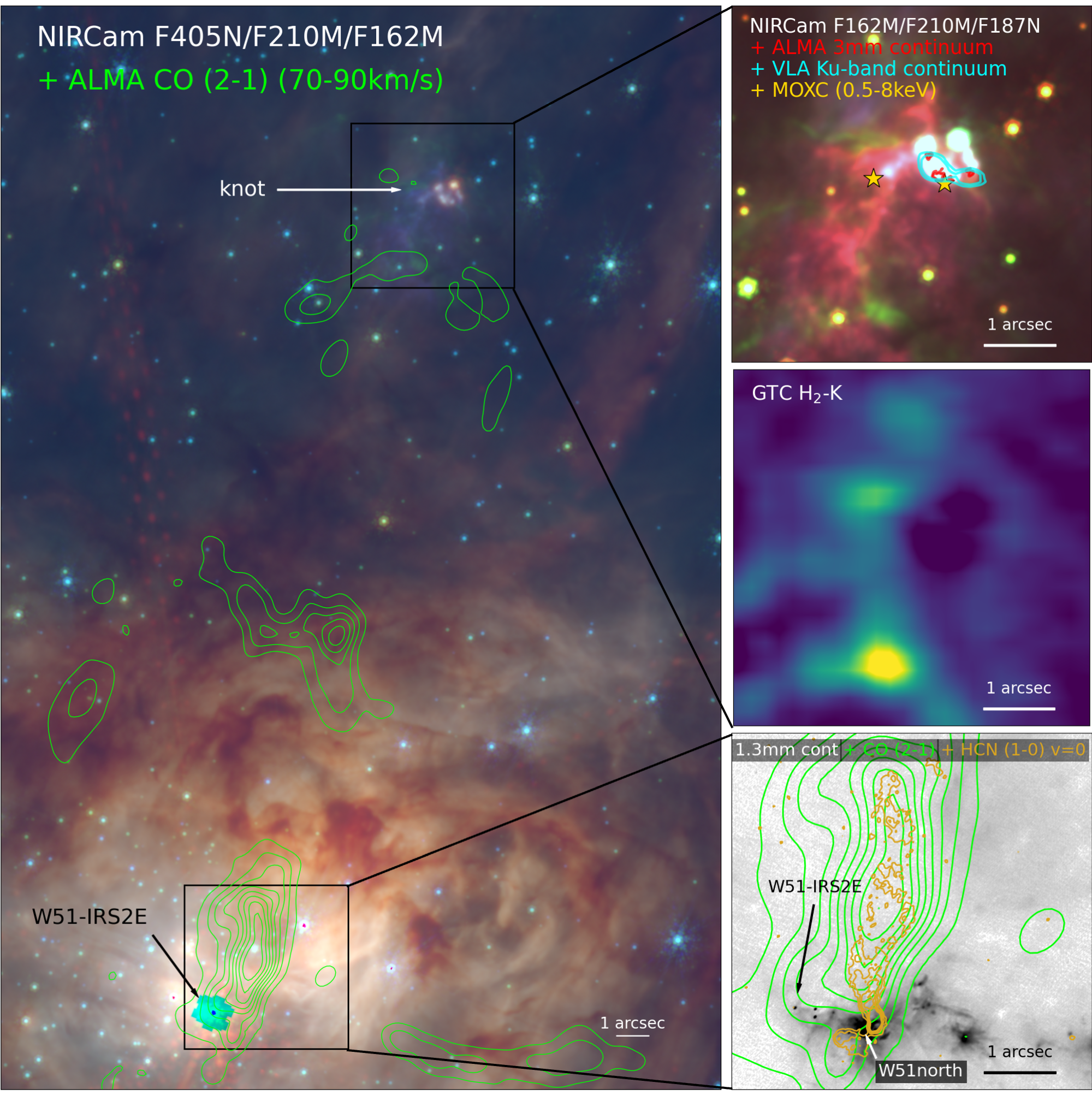}
    \caption{Outflow structure around  W51-IRS2. The orientation of all panels is aligned with ICRS coordinates. \textbf{Left}: The background image is a composite of NIRCam F162M (red), F210M (green), and F405N (blue). The green contour shows the ALMA CO (2-1) moment 0 map from the ALMA-IMF large program \citep{motte22, cunningham23} with a radial velocity range of 70-90 km/s. \textbf{Upper right}: A zoomed-in composite of F162M (red), F210M (green), and F187N (blue) for the outflow knot. The red contour displays the ALMA $3\,{\rm mm}$ continuum above $3\sigma=5.25\,{\rm mJy/beam}$ from ALMA observation 2017.1.00293.S \citep{ginsburg19}. The cyan contour exhibits the VLA Ku-band continuum. \textbf{Middle right}: GTC \htwo\ -K image \citep{dawson25} for the knot cutout. \textbf{Lower right}: Another zoomed-in image around W51-IRS2E and W51north. The background image is an ALMA $1.3\,{\rm mm}$ long baseline (2015.1.01596.S; \citealt{ginsburg19}) image created by feathering with the ALMA-IMF image. ALMA CO (2-1) (green contour) and HCN (1-0) v=0 (yellow contour) moment maps are displayed as well. }
    \label{fig:outflow}
\end{figure*}

We identified a knot at the position of an outflow-associated shock structure previously reported in the literature \citep{hodapp02, ginsburg16, dawson25}. This knot is spatially coincident with the bow shock morphology seen in the \htwo\ S(1) image of \citet{hodapp02} and in the \htwo-K image from the GTC observations (\citealt{dawson25}; D25H2 \#1).
%The suggested bow shock can be reconfirmed in  Fig.~\ref{fig:outflow} B exhibiting several wavy structures of molecular hydrogen at F210M (green) and diffuse emission of [Fe II] (red).
Furthermore, two Chandra observation X-ray sources from MOXC (IDs: CXOU 192339.69+143130.4 and 192339.62+143130.3; \citealt{townsley14}) were observed near the knot.

This knot system is surprisingly bright, especially at long wavelengths.
The rising SED from 1--$4\,{\rm \mu m}$, combined with the detection in ALMA continuum and the point-like morphology of the sources would naively be best interpreted as a YSO, i.e., a stellar photosphere heating a massive disk to provide the infrared excess.
On the other hand, the morphology of this feature strongly suggests that it is a bow shock, and there is additional new evidence that more strongly favors this scenario.
The VLA Ku-band / ALMA 3mm spectral index integrated over the whole feature is about 0.3, which indicates that the emission is dominated by free-free radiation.
% only be more quantitative about this if we absolutely have to
There is a strong recombination line excess seen in F405N and F187N, as shown in Fig.~\ref{fig:outflow}.
While this excess holds for the full area, we note that there is no mm or cm continuum detected in the uppermost (northernmost in ICRS) knot, and it shows a stronger mid-infrared excess than the others. \cite{ginsburg16} suggested that the VLA continuum sources d4 at the location of the knot could be created by the interaction between the outflow and the ISM. 
In particular, the d4 sources were reported to be a variable in $\sim20\,{\rm year}$ timescale, further suggesting the products of the interaction of shock transients \citep{ginsburg16}. 

% speculative
It is possible that the upper side of the feature shows the forward bow shock, and the lower side, with mm and cm continuum detections, shows the reverse shock. However, it is not obvious why the reverse shock should be more luminous at long wavelengths. On the other hand, the hypothesis that the knot is a YSO or protostellar system interacting with the outflow cannot be completely ruled out. For instance, the recent study of the high-mass protostar GGD 27-MM2(E) found an X-shaped shocked region around the molecular core, which is thought to be created by the interaction between the molecular core and the outflow on $\sim7000\,{\rm AU}$ scale \citep{lopez-vazquez25}.
The projected size of the bow shock ($\sim 10^4$ AU) relative to the length of the outflow ($\sim 0.7$ pc) gives a sky filling area of about 0.5\%,  such that a chance impact of an outflow on a YSO is quite low but not negligible.

%Given the projected size of the outflow in W51-IRS2, $r\sim0.7\,{\rm pc}$, however, the chances of the outflow hitting a compact source are very low.

% I think the English here conveys the wrong message, so I'm rewriting it.
The knot and outflow structure found is similar to the HH80/81 shock, which also appears to be interacting with a dense ISM structure at its endpoint \citep[e.g.][]{bally23}.
Both HH80/81 and D25H2 \#1 are, unlike the majority of Herbig-Haro shocks, detected in X-rays.
The MOXC sources associated with D25H2 \#1 have total X-ray flux $10^{-15}\,{\rm mW/m^2}$ at $0.5$--$8\,{\rm keV}$, which is comparable to the X-ray flux found in HH80/81, which is $S_{5.4 \mathrm{~kpc}}=7\times10^{-16}\,{\rm mW/m^2}$ at the distance of W51, $d=5.4\,{\rm kpc}$ \citep{lopez-santiago13}.
If we assume D25H2 \#1 has a similar velocity to HH80/81, $\sim1000\,{\rm km/s}$ \citep{bally23}, the proper motion is approximately $1\,{\rm mas/yr}$, which will be measurable using either ALMA or JWST observations over $>10$-year baseline.

\cite{hodapp02} pointed out that the driving source of the outflow is likely W51-IRS2, but we can now pinpoint its origin more precisely. The CO (2-1) emission map from ALMA-IMF ALMA large program \citep{motte22, ginsburg22, cunningham23} and HCN (1-0) v=0 emission from ALMA long-baseline observations \citep{yoo25} connect the knot to the driving source, W51north (lower right panel of Fig.~\ref{fig:outflow}). W51north is a massive protostar with a mass of $14$--$28\,M_\odot$ \citep{goddi20}.
The H$_2$ arc morphology supports this interpretation: 
there are several H$_2$ arcs (seen in the F210M filter; green in Figure \ref{fig:outflow}) with corresponding detection in ALMA CO 2-1 emission at $70\,{\rm km/s} < v < 90\,{\rm km/s}$, providing kinematic evidence that the outflow driven from W51 North is the driver of this bow shock.
%The F405N image in Fig.~\ref{fig:outflow} shows two large arcs launched from W51-IRS2 that are nearly parallel to, but not exactly coincident with, the outflow vector. These arcs may circumscribe an outflow cavity, but because the knot is off-center by $\sim1\,{\rm arcsec}$, it is unclear if it is this outflow or another.
% not relevant here: Moreover, \cite{goddi20} investigated SiO J=5-4 outflow from W51north and concluded that the old SiO outflow direction is consistent with the CO (2-1) outflow, while the recent young SiO flow is tilted by $\sim70\,{\rm deg}$. 

\section{Summary}

In this paper, we present the first look at JWST NIRCam/MIRI images of the W51A region. With sub-arcsecond angular resolution, the JWST images give detailed morphological features of dust filaments and H\textsc{ii} regions. Using NIRCam/MIRI multi-filters, we separate the region with warm dust (F480M-F360M), PAH abundance (F335M/F480M), and ionized gas (F405N-F410M) components. We find that F335M/F480M is spatially anti-correlated with F480M-F360M and F405N-F410M, showing that PAH is mainly destroyed in the ionized region. We compare our JWST images with other multiwavelength data such as the ALMA continuum images and the GTC \htwo-K images, to explore their high-resolution infrared views. In particular, we find nearly 11\% of matches among $\sim200$ of the total ALMA compact sources in the region, which suggests that still a significant fraction of the ALMA-detected protostellar population has high dust extinction due to either their dust envelope or the dust content along the line-of-sight. Among the ALMA-matched sources, we report peculiar sources, W51-IRS2E and a knot in W51-IRS2. W51-IRS2E is particularly bright at wavelength $\lambda\gtrsim2\,{\mu m}$, saturating a broad region around W51-IRS2. Its unique spatial association with hard X-ray emission and CO overtone emission, combined with the compact nature of its ALMA counterpart, makes this source particularly worthy of further investigation. The exceptionally IR-bright knot in W51-IRS2 is spatially associated with X-ray emission, bow shock features of CO (2-1) emission and \htwo\ emission, 
%JH
which 
favors the scenario where the knot is created by the interaction between the outflow and the ISM. The trail of CO (2-1) emission points toward another massive protostar in W51-IRS2, W51north, with an estimated mass of $14$--$28\,M_\odot$.

We defer the detailed characterization of the emission we observe in NIRCAM and MIRI to future papers. To this end, the spectral energy distribution should be obtained with photometry or spectroscopy.  

\begin{acknowledgements}
    This work is based on observations made with the NASA/ESA/CSA James Webb Space Telescope. The data were obtained from the Mikulski Archive for Space Telescopes at the Space Telescope Science Institute, which is operated by the Association of Universities for Research in Astronomy, Inc., under NASA contract NAS 5-03127 for JWST. These observations are associated with program \#6151. Support for program \#6151 was provided by NASA through a grant from the Space Telescope Science Institute, which is operated by the Association of Universities for Research in Astronomy, Inc., under NASA contract NAS 5-03127.
    
    AG and TY acknowledge support from the Space Telescope Science Institute via grant No. JWST-GO-06151.001-A. and from NSF under grant CAREER 2142300.
    NB acknowledges support from the Space Telescope Science Institute via grant No. JWST-GO-05365.001-A. RGM acknowledges support from  UNAM-PAPIIT project IN105225. J.S-B acknowledges the support received by the UNAM DGAPA-PAPIIT project AG-101025 and from the SECIHTI Ciencia de Frontera project CBF-2025-I-3033. MGSM thank the Spanish MCINN for funding support under grant PID2023-146667NB-I00 funded by MCIN/AEI/10.13039/501100011033. 
    
    The authors acknowledge University of Florida Research Computing for providing computational resources and support that have contributed to the research results reported in this publication. URL: http://www.rc.ufl.edu.

\end{acknowledgements}
\begin{contribution}
TY obtained the data, led the project, and wrote the manuscript. AG obtained the data, supervised the progress of the research, and revised the writing. NB and RGM contributed to interpreting the analyzed data and revised the writing. AD provided the GTC data for comparison. SG, JH, ARL, CGRZ, JSB, MGSM, AW, and JEY contributed to the revision of the manuscript.
\end{contribution}
\facilities{JWST, ALMA}
\software{JWST standard pipeline version 1.17.1 \citep{jwst_pipeline_1.17.1}, \texttt{astropy} \citep{astropy13,astropy18}}
\bibliography{sample701}{}
\bibliographystyle{aasjournalv7}

\end{document}